\documentclass[usegraphicx,useAMS,usenatbib]{mn2e}
\usepackage{times}
\usepackage{amssymb}
\usepackage{amsmath}
\usepackage{graphicx}
\usepackage{euscript}
\usepackage{rotating}
\usepackage{epsfig}
\usepackage{mathptmx}
\usepackage{amsbsy}
\DeclareMathAlphabet{\vib}{OML}{cmm}{m}{it}
\begin{document}
\include{defn}
\title{Detections of Molecular Hydrogen in the Outer Filaments of NGC\,1275}
\author[N. A. Hatch, C. S. Crawford, A. C. Fabian \& R. M. Johnstone]
{N. A. Hatch,\thanks{E-mail:nah@ast.cam.ac.uk} C. S. Crawford, A. C. Fabian and R. M. Johnstone\\
Institute of Astronomy, Madingley Road, Cambridge, CB3 0HA}
\maketitle
\begin{abstract}
We present clear spectroscopic detections of molecular hydrogen in the outer filaments of the H$\alpha$ nebula surrounding the central galaxy of the Perseus cluster, NGC 1275. This implies the presence of warm molecular gas clouds at projected distances up to 24\,kpc from the nucleus of the host galaxy, embedded in the hot intracluster medium. The emission-line intensity ratios reveal that the H$_{2}$ emission is predominately thermally excited with excitation temperatures of 1600$-$2200\,K, suggesting a lack of pressure balance between the molecular component and its surrounding medium. Excitation by stellar UV or by the central AGN is shown to be unlikely, whilst thermal excitation by X-rays or conduction from the ICM or shocks are all possibilities. Evidence for a non-thermal component is found in the spectra of some regions based on a low ortho-to-para ratio.
\end{abstract}
\section{Introduction}
Luminous optical and UV line-emitting filamentary nebulae like that of NGC 1275 are commonly found surrounding central cluster galaxies where the X-ray emitting intracluster medium (ICM) has a short radiative cooling time (\citealt{Crawford} and references therein). Despite many studies, the origin and excitation mechanism of these filaments are uncertain, with cooling of the intracluster gas, star formation, conduction and shock heating all having been suggested and contested (e.g. \citealt{RodAndy}, \citealt{SaraShieldsFlip}).

The large H$\alpha$ nebulousity surrounding NGC 1275 has been long known (\citealt{Minkowski}, \citealt{Lynds}) and has been studied in a variety of wavelengths. Deep H$\alpha$ and X-ray images obtained by the WIYN telescope \citep{conselice} and Chandra \citep{Fabian2003} show spectacular, extended H$\alpha$ filamentary structures stretching over 50\,kpc from the central galaxy, with corresponding soft X-ray counterparts, deeply embedded in the hot ICM.

NGC 1275 lies at the centre of a cluster with strong centrally peaked X-ray emission where the X-ray luminosity implies a mass deposition rate of 300\,M$_{\odot}$yr$^{-1}$ \citep{Allen}. However, after extensive searching, the massive amounts of gas expected to accumulate cannot be accounted for as stars or line-emitting gas near the nucleus. 

Chandra spectra show little evidence of gas cooling below a temperature about one third of that in the bulk of the cluster \citep{Schmidt,Sanders}. Therefore most of the gas either cools further in a non-radiative manner, requiring energy to be shared between the hot X-ray emitting and cooler gas by mixing or conduction, or is heated by a distributed heat source. A number of theories including heating by the central active galactic nucleus (AGN) of NGC 1275 have been suggested (see \citealt{Fabian2004} and references therein).

A reservoir of $\sim$4$\times$10$^{6}$\,M$_{\odot}$ \citep{Heckman} of 10$^{4}-10^{5}$\,K gas lies in the optical and UV line-emitting nebula. Molecular gas appears common in such regions \citep{Jaffe97}. Recent observations show that there is a component of warm (1000$-$4000\,K) molecular hydrogen gas \citep{Donahue,Edge} with up to $\sim$1$\times$10$^{10}$\,M$_{\odot}$ of cool molecular gas \citep{Edge2}. \citet{Krabbe} and \citet{Donahue} deduced that the molecular hydrogen component in the NGC 1275 nuclear region is likely to be excited by the central AGN, whilst \citet{Wilman} argue that the H$_{2}$ emission is thermal and mostly likely to be heated by UV radiation. The outer filaments have a fairly constant H$\alpha$ surface brightness extending over many tens of kpc, therefore a central excitation source such as the AGN is unlikely to be of great importance. Whilst massive stellar clusters are seen in H$\alpha$, they do not spatially correspond with all the filaments and bright knots. Thus the excitation of the outer filaments may be different to that of the nuclear region.

The redshift of NGC 1275 is 0.0176, which using H$_{0}$=70\,kms$^{-1}$Mpc$^{-1}$, gives 1\,kpc$\simeq$2.7\,arcsec.
\section{Observations}
\begin{table*}
\begin{tabular}{|lllrrc|}
Target &\multicolumn{2}l{Slit centre co-ordinates} &Position  &Nodding&Total integration \\ 
&RA & Dec (J2000)&angle& (arcsec)& time on target (minutes)\\ \hline
Slit 1 (NW) & 03 19 46.78&+41 31 45.28  &$-$125 $\mathring{}$&0,  $-$10&36\\
Slit 2 (SW)&03 19 41.95& +41 29 58.61&$-$76 $\mathring{}$ &$-$30,  $-$20&48\\
Slit 3 (E)& 03 19 49.31& +41 31 50.52& $-$4 $\mathring{}$ &$-$35,  $-$80&54\\ \hline
\end{tabular}
\caption{ Log of Observations. Nodding $^{+}_{-}$ direction and position angle as explained in CGS4 literature. }
\label{slits}
\end{table*}
The spectra presented in this paper were taken with the CGS4 spectrograph on the United Kingdom Infrared Telescope on 2004 January 17, 18 and 19 for a total integration time of 15 hours. The 40\,lines mm$^{-1}$ grating was used with the 300\,mm focal length camera, giving a spatial scale of 0.61\,arcsec per pixel. The B2 filter centred at 2.02\,$\micron$ was used with a two pixel wide slit giving a spectral resolution of 600$-$790\,kms$^{-1}$. The observations were taken in NDSTARE mode, using a object-sky-sky-object nodding pattern. The seeing was better than one arcsec on all three nights and there was little to no cloud cover.

 The telescope was directed to a bright reference source near the target position. The slit was then shifted to a targeted bright object (co-ordinates given in Table \ref{slits}) and orientated to the correct position angle. For confirmation of the exact spatial position of the slit we included one or two bright objects such as a star or galaxy along the length of the slit. Table \ref{slits} gives the co-ordinates, position angle and integration time for all observations. Guided by the H$\alpha$ map of \citet{conselice}, the slit positions were chosen to include some of the brightest knots and filaments furthest from the nucleus. 

Most sky emission features were removed automatically by the nodding pattern. An atmospheric, temporally variable CO$_2$ line near 2.01\,$\micron$ and a feature at 2.15\,$\micron$ however remain in the spectrum from the Slit 1 and 2 position. The target observations were preceded and followed by a series of calibration observations including flat-fields, dark frames, arcs and a set of standard F and G type stars. These standards were nearly featureless stars of known magnitude. In each of the standard stars a Br$\gamma$ stellar feature at 2.166\,$\micron$ was removed by linear interpolation. The data was reduced with ORAC-DR Version V4.0-1 available through the UKIRT software website. Data manipulation and extraction was done with Starlink packages, QDP \citep{qdp} and IDL Version 6.0.
\section{Results}
\begin{figure*}
 \centering
\includegraphics[width=1.099\columnwidth]{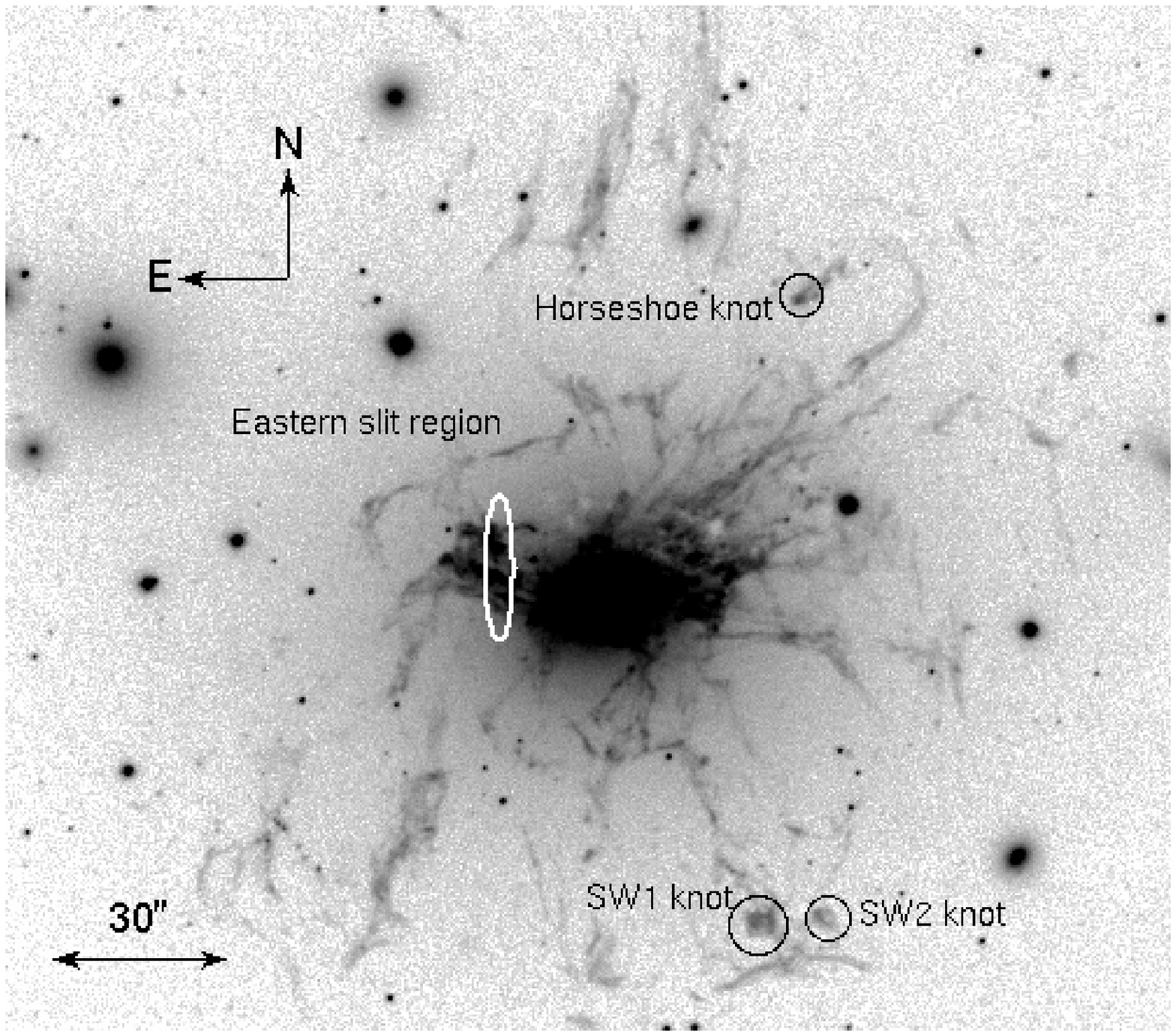}
\includegraphics[width=0.901\columnwidth]{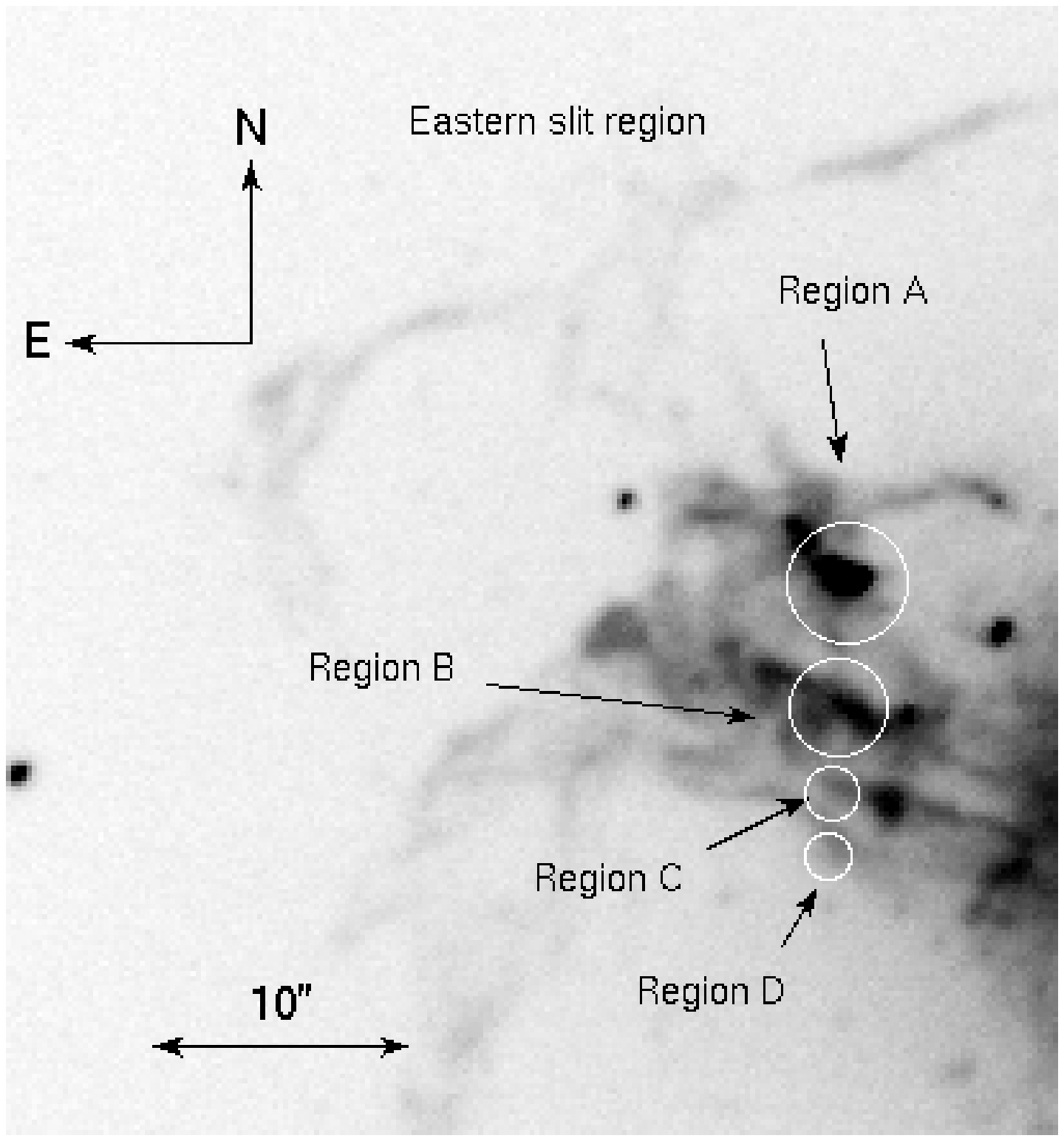}
\caption{Left: WIYN telescope H$\alpha$ image of the line-emitting filamentary structure surrounding NGC1275 from \citet{conselice}. Regions where H$_{2}$ ro-vibrational lines are detected have been marked by circles and labelled. Right: Detail of the Eastern slit region.\label{perreg}}
\end{figure*}
\begin{figure*}
\centering
\includegraphics[width=1.45\columnwidth, angle=-90]{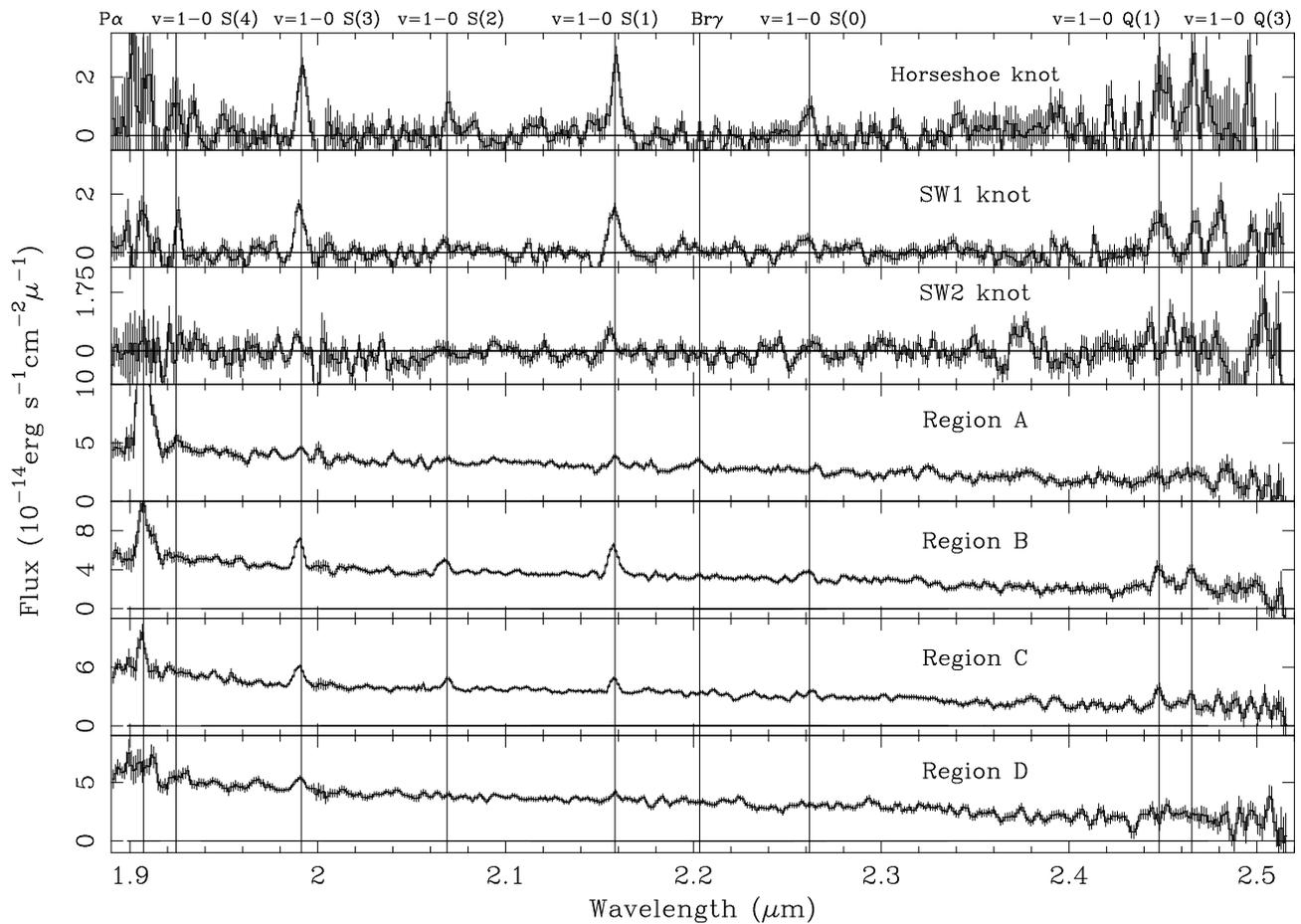}
\caption{UKIRT K-band smoothed spectra of the outer filament regions marked in Fig. \ref{perreg} with the ro-vibrational H$_{2}$ and atomic hydrogen lines labelled. Line labels have redshift z=0.0172. Errors are from Poisson Statistics. Note that the scale in Region A does not show full extent of P$\alpha$.}
\label{Spectra}
\end{figure*}

Clear detections of the strongest molecular hydrogen lines: $\vib{v}$=1$-0~$S(1) and $\vib{v}$=1$-$0 S(3) were found in seven separate regions of the outer filaments, marked and labelled in Fig. \ref{perreg}. Region A covers an area of very strong H$\alpha$ emission. \citet{ShieldsFlip} discovered and studied a massive, young stellar cluster in this region. It lies 19.4\,arcsec (a projected distance of 7.2\,kpc) to the East of the central galaxy, marked as a bright knot in H$\alpha$ within the Northern bright filament that is part of array of radial filaments that point East from the galaxy. Regions B, C and D come from this array of radial filaments at distances 18.0\,arcsec (6.7\,kpc), 17.7\,arcsec (6.6\,kpc), 17.9\,arcsec (6.6\,kpc)  from the central galaxy respectively. There is no clear distinction separating regions C and D in the slit image, although spectral analysis revealed they were most likely separate populations of molecular hydrogen. The {\textquoteleft horseshoe knot\textquoteright}  lies at the tip of a horseshoe-shaped feature at the end of a 74\,arcsec long NW radial filament that appears to turn back on itself, noted by \citet{conselice} (box 11 in their Fig. 5). It has a projected distance of 22\,kpc away from the galaxy center. SW1 (box 2 in \citealt{conselice} Fig. 5) and SW2 are bright knots that lie at the end of radial filaments (61.6\,arcsec $\sim$23\,kpc and  64.6\,arcsec $\sim$24\,kpc from the galaxy respectively) pointing SSW. 
\begin{table}
\centering
  \begin{tabular}{|l|l|r|}\hline
Region&Line&Surface Brightness  \\
RA$~~~~~~~~~~~~~~~~~~~~~$Dec (J2000)&&\\
size of extracted region&&\\\hline
Horseshoe knot&$\vib{v}$=1$-$0 S(3)  &20.7 $^{+3.6}_{-3.3}$\\
03 19 45.22$~~~~$ +41 31 32.61 &  $\vib{v}$=1$-$0 S(2) & 6.6 $^{+2.8}_{-2.5}$\\
2.23 arcsec$^{2}$ &$\vib{v}$=1$-$0 S(1) &  18.1 $^{+3.2}_{ -2.8}$\\
 &$\vib{v}$=1$-$0 S(0)  & 5.2 $^{+3.6}_{-2.1}$\\ \hline
SW1 knot &P$\alpha$  &11.5$^{+7.2}_{-4.1}$\\  
03 19 45.69$~~~~$ +41 29 46.99 &$\vib{v}$=1$-$0 S(4) & 7.5 $^{+2.9}_{-2.7}$\\
4.47 arcsec$^{2}$&$\vib{v}$=1$-$0 S(3) & 13.1 $^{+1.8}_{ -1.7}$\\
 &$\vib{v}$=1$-$0 S(2) & 2.6 $\pm1.4$\\
 &$\vib{v}$=1$-$0 S(1) & 15.3 $^{+2.3}_{-2.1}$\\ 
&$\vib{v}$=1$-$0 S(0) & 4.9 $^{+1.9}_{-1.8}$\\ 
&$\vib{v}$=1$-$0 Q(1) & 14.7 $^{+4.3}_{-4.2}$\\\hline
SW2 knot&$\vib{v}$=1$-$0 S(3) & 4.6 $^{+2.7}_{-2.2}$\\
03 19 44.97$~~~~$ +41 29 48.92  &  $\vib{v}$=1$-$0 S(1) & 5.6 $^{+2.1}_{-1.9}$\\ 
2.98 arcsec$^{2}$&&\\ \hline
Region A& P$\alpha$(Narrow)  &119.1 $\pm$35.6\\ 
03 19 49.71$~~~~$ +41 30 51.11&P$\alpha$(Broad)  &38.7 $\pm$35.6\\
2.23 arcsec$^{2}$  &$\vib{v}$=1$-$0 S(3)&4.6 $^{+1.3}_{ -1.4}$\\
&$\vib{v}$=1$-$0 S(1) &6.5 $^{+1.7}_{-1.9}$\\
& Br$\gamma$&8.3 $^{+5.6}_{-2.6}$\\ \hline
Region B&P$\alpha$  &45.1$^{+10.9}_{-9.0}$\\  
03 19 49.75$~~~~$ +41 30 45.19&$\vib{v}$=1$-$0 S(3)&23.5 $^{+2.6}_{-2.5}$\\
2.23 arcsec$^{2}$ &$\vib{v}$=1$-$0 S(2)  &12.5 $^{+2.6}_{-2.4}$\\ 
 &$\vib{v}$=1$-$0 S(1)   &24.5 $^{+2.5}_{-2.3}$\\ 
 &$\vib{v}$=1$-$0 S(0)  &9.8 $^{+2.7}_{-2.4}$\\
 &$\vib{v}$=1$-$0 Q(1) &17.2 $^{+5.3}_{-4.9}$\\
 &$\vib{v}$=1$-$0 Q(3) &18.6 $^{+8.5}_{-6.5}$\\ \hline
Region C&P$\alpha$ &25.5 $^{+6.4}_{-6.1}$\\
 03 19 49.77$~~~~$ +41 30 43.24&$\vib{v}$=1$-$0 S(3)  &16.8 $\pm$1.9\\
2.98 arcsec$^{2}$ &$\vib{v}$=1$-$0 S(2)  &7.3 $^{+1.9}_{-1.7}$\\
 &$\vib{v}$=1$-$0 S(1)  &10.5 $\pm$1.6\\
 &$\vib{v}$=1$-$0 S(0)  &4.6 $^{+1.7}_{-1.6}$\\
 &$\vib{v}$=1$-$0 Q(1)  &10.2 $^{+4.4}_{-4.2}$\\ \hline
Region D &$\vib{v}$=1$-$0 S(3) &10.3 $^{+3.0}_{-2.5}$\\
03 19 49.77$~~~~$ +41 30 41.28  &$\vib{v}$=1$-$0 S(1) &5.1 $^{+1.5}_{-0.4}$\\ 
2.23 arcsec$^{2}$ &&\\ \hline
\end{tabular}
\caption{\label{li} Line surface brightness in $10^{-17}$erg s$^{-1}$cm$^{-2}$arcsec$^{-2}$ for all regions (as marked in Fig. \ref{perreg}) in which the strongest ro-vibrational H$_{2}$ lines, $\vib{v}$=1$-$0 S(1) and $\vib{v}$=1$-$0 S(3) are clearly detected. Errors are at 1\,$\sigma$ level.}
\end{table}
Our K-band spectra for all regions are presented in Fig. \ref{Spectra}, with line surface brightness given in Table \ref{li}. Since the structures observed may be smaller than the slit size, these intensities may be underestimated, hence they could be considered lower limits. The spectra have not had any stellar continuum removed. The SW regions and the horseshoe knot show no evidence for a continuum, whereas regions A, B, C and D all have a continuum due to the galactic stellar emission. 
 
In principle the wavelength range should allow an estimation of the extinction using the P$\alpha$ and Br$\gamma$ lines or from the $\vib{v}$=1$-0~$Q(3) and $\vib{v}$=1$-0~$S(1) lines. However the P$\alpha$ and $\vib{v}$=1$-0~$Q(3) emission lines pass through complex parts of the atmospheric window and the Br$\gamma$ emission is too faint to have a significant detection in all but one region. Therefore no reddening has been taken into account in this work, although we note any consequence it may have in our discussion of the data.

\section{Analysis and Discussion}
\subsection{Velocity Structure}

\begin{figure}
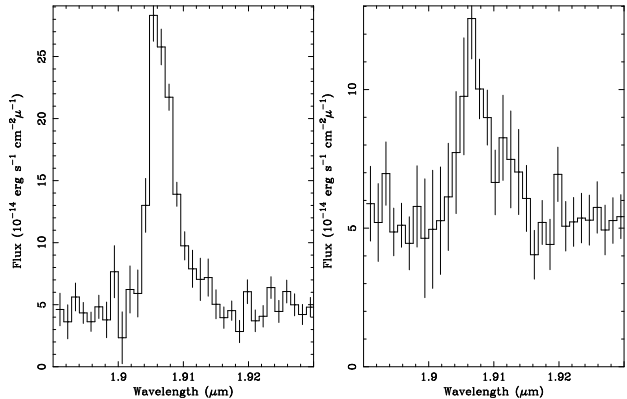

\centering
\includegraphics[width=0.62\columnwidth , angle=-90]{fig3a.ps}
\includegraphics[width=0.62\columnwidth , angle=-90]{fig3b.ps}
\caption{P$\alpha$  line emission from region A,  the young stellar cluster (left) and region B (right), both showing non-Gaussian profiles.}
\label{pa}
\end{figure}

The P$\alpha$ emission from regions A and B both exhibit non-Gaussian profiles, shown in Fig. \ref{pa}.  This profile could not occur through line blending as the P$\alpha$ line is distinct from the other molecular and atomic hydrogen lines. These profiles may be the result of either extra redshifted material or absorption of the blueshifted emission, either intrinsically or by the Earth's atmosphere (this line is located in a complex part of the atmospheric window). Neither the molecular hydrogen lines nor the other atomic lines from our spectra exhibit these profiles, although given the high P$\alpha$/H$_{2}$ line ratio of region A, it is unlikely that a low flux red wing would be visible in the molecular emission above the noise level. 

The H$\alpha$ emission from region A has been studied by \citet{Shields2} who find the H$\alpha$ profile shows a well defined narrow peak at z=0.0168 and a large bump of emission at a larger redshift. They interpret the lower redshift, narrow emission as originating from the stellar cluster, and the broad emission originating from overlapping filamentary material. From the [O{\sc{i}}]/H$\alpha$ ratio they estimated approximately 30 per cent of the total H$\alpha$ luminosity in this region originates from the  filament. The P$\alpha$ emission profile from region A can be better fitted by two Gaussians, a narrow Gaussian at z$\sim$0.0165 containing 75 per cent of the total line emission and a broad Gaussian containing 25 per cent of the line emission with z$\sim$0.018, rather than a single Gaussian. Although our slit may not have covered the exact same region investigated by \citet{Shields2}, the similarities in line profile suggest the same interpretation.

Region B can also be fitted by two Gaussians; a broad dominating component at z$\sim$0.0169 and a narrow component  at a much larger redshift. The large errors and lack of other corroborating evidence makes it difficult to apply an intrinsic emission interpretation to this region as opposed to absorption by the atmosphere. Therefore a single Gaussian profile is fitted to the P$\alpha$ emission. 

\begin{figure}
\centering
\includegraphics[width=1\columnwidth ]{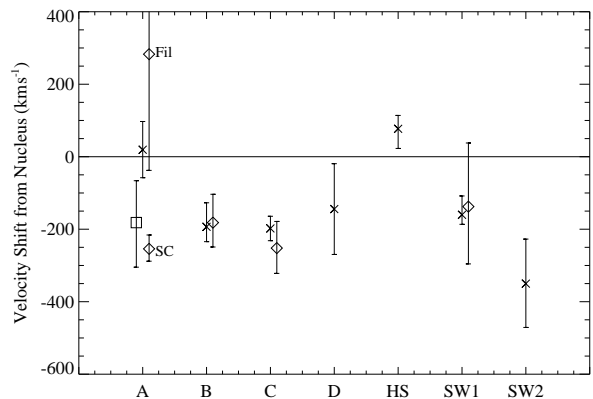}
\caption{Velocity shift of molecular and atomic hydrogen lines relative to the nucleus, measured from a combined fit of the $\vib{v}$=1$-$0 S(1), $\vib{v}$=1$-$0 S(2) and $\vib{v}$=1$-$0 S(3) lines (crosses), P$\alpha$ lines (diamonds), Br$\gamma$ (square). HS is the horseshoe knot. The velocity shift of both the narrow P$\alpha$ emission labelled A(SC) and the broad emission labelled A(fil) are shown. Velocity resolution is approximately 700\,kms$^{-1}$, errors reflect 1\,$\sigma$ level on line position. Nuclear molecular hydrogen line positions from \citet{Krabbe}; nuclear P$\alpha$ (1.8756\,$\micron$) and Br$\gamma$ (2.1661\,$\micron$) line positions chosen to conform with redshift of z=0.0172 as measured by \citet{Krabbe}}
\label{vel}
\end{figure}

None of the lines were resolved spectrally but Fig. \ref{vel} shows that there is a velocity shift relative to the nucleus of the strong H$_{2}$ lines: $\vib{v}$=1$-$0 S(1), $\vib{v}$=1$-$0 S(2), $\vib{v}$=1$-$0 S(3) and P$\alpha$ emission.  This figure shows the molecular and atomic emission from regions B, C and the SW1 knot appear to have similar velocity shifts, therefore it is likely they share an origin. The Br$\gamma$ emission from region A has a similar velocity shift to the narrow P$\alpha$ emission, therefore most likely to have originated from the same region, the stellar cluster. The molecular emission from region A does not share the same velocity shift as either the stellar cluster or the filament. It is possible the molecular hydrogen emission is a superposition of emission from both the filament and the stellar cluster, or the molecular emission may not share the same origin as the atomic emission. If the P$\alpha$ emission was fit with a single profile, the velocity shifts of the P$\alpha$ and the H$_{2}$ would still not agree. This figure suggests that the outer filaments observed are systematically blueshifted relative to the nucleus.

\subsection{H$_{2}$ Excitation}
\begin{figure}
\centering
\includegraphics[width=1\columnwidth]{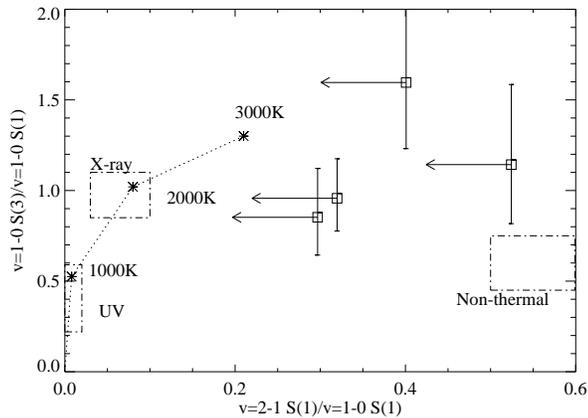}
\caption{Diagram constraining likely excitation mechanisms. The four 3$\sigma$ upper limits indicate observed line ratios from the brightest filaments, errors on the $\vib{v}$=1$-$0 S(1)/$\vib{v}$=1$-$0 S(3) ratio are 1$\sigma$ level. Boxes highlight regions occupied by the non-thermal UV excitation models of \citet{BlackDishoeck} and models of thermal processes near the dotted thermal curve including UV excitation \citep{SternbergDaldgarno} and X-ray heating models (\citealt{LeppMcCray} and \citealt{DraineWoods}). The x-axis gives an indication of the level of thermal to non-thermal excitation occurring. The upper limits are consistent with a thermal excitation mechanism, but there also may be a non-thermal component.}
\label{v2-1}
\end{figure}
 H$_{2}$ excitation may occur through collisions and therefore exhibit thermal emission, or from radiative transfers producing non-thermal emission. These mechanisms can be distinguished since they preferentially populate different levels producing different spectra, but as they often occur together the observed spectra can exhibit both thermal and non-thermal emission. Collisional de-excitation dominates when the gas has a total density n$_{{\rm T}}=$ n$_{{\rm H}}+$n$_{{\rm H_{2}}} \ge 10^{4}$\,cm$^{-3}$ \citep{SternbergDaldgarno,Mandy} for T$\sim$10$^{3}$\,K. The gas may be heated by shocks, conduction from the ICM, X-rays or UV photons and the H$_{2}$ energy levels populated through collisional transitions. When this occurs the ortho-to-para ratio is three and the gas is in local thermodynamic equilibrium (LTE). Below the critical density, the probability of the ${\rm H_{2}}$ undergoing a transition from an upper to a lower energy level via a radiative process is greater than for collisional de-excitation. The H$_{2}$ molecules can be excited to the upper energy levels through non-thermal excitation mechanisms such as X-rays and UV photons in the Lyman-Werner band (912$-$1108$\mathring{{\rm A}}$) or through collisions with suprathermal electrons caused by X-ray ionization.

Non-thermal excitation mechanisms readily excites the $\vib{v}=$ 2 and higher vibrational states, whereby collisional transitions preferentially de-excite the $\vib{v}=$ 2 level in favour of the $\vib{v}=$ 1 level, so the $\vib{v}$=2$-$1 S(1)/ $\vib{v}$=1$-$0 S(1) ratio is an indicator of the relative level of thermal to non-thermal processes. Reddening does not have a large effect in this part of the spectrum, therefore it can be ignored without much consequence when considering H$_{2}$ line ratios. Models predict a $\vib{v}$=2$-$1 S(1)/$\vib{v}$=1$-$0 S(1) ratio greater than 0.53 for pure fluorescence at n$_{T} \le$10$^{5}$\,cm$^{-3}$ and low temperatures \citep{DraineBert}, whilst shock models of \citet{Hollenbach} exhibit ratios between 0.04$-$0.46. 3$\sigma$ upper limits of the $\vib{v}$=2$-$1 S(1) emission from the four regions exhibiting the strongest H$_{2}$ lines and best signal-to-noise ratio (region A, region B, SW2, horseshoe knot) are plotted in an excitation diagram (Fig. \ref{v2-1}). All lines used in these ratios are ortho, so this diagram is independent of an uncertain ortho-to-para ratio.
 Fig. \ref{v2-1} is consistent with predominately thermal (collisional) excitation occurring but does not rule out a non-thermal component.

If collisional de-excitation is assumed to dominate, the H$_{2}$ will be in local thermal equilibrium (LTE). The energy levels of H$_{2}$ will be populated in a Boltzmann distribution and from this an excitation temperature (T$_{{\rm ex}}$) can be derived through
\begin{equation}\nonumber 
\frac{N(\vib{v}, J)}{g_{J}}= a~e^{-E(\vib{v}, J)/{\rm k_{b}}T_{{\rm ex}}}
\end{equation}
where $E(\vib{v}, J)$ is the upper energy of the $(\vib{v}, J)$ transition, $a$ is a constant, $\rm{k_{b}}$ is Boltzmann's constant and g$_{J}$ is the statistical weight. $N(\vib{v}, J)$ is the transition column density, which, for optically-thin emission is given by
\begin{equation}\nonumber
N(\vib{v}, J)=\frac{4 \pi I}{{\rm A_{ul}} h\nu}
\end{equation}
where $I$ is the observed intensity of the line, $\nu$ is the rest frequency and  A$_{\rm{ul}}$ is the Einstein co-efficient taken from \citet{Turner}. In all calculations an ortho-to-para ratio of 3:1 has been assumed. If the H$_{2}$ is only thermally excited, the excitation temperature is the kinetic temperature. However, if non-thermal excitation is also taking place, the excitation temperature is not the actual temperature of the gas, but characterises the level populations that have arisen through competing thermal and non-thermal processes \citep{Puxley}.

\begin{figure*}
\centering
\includegraphics[width=1.8\columnwidth]{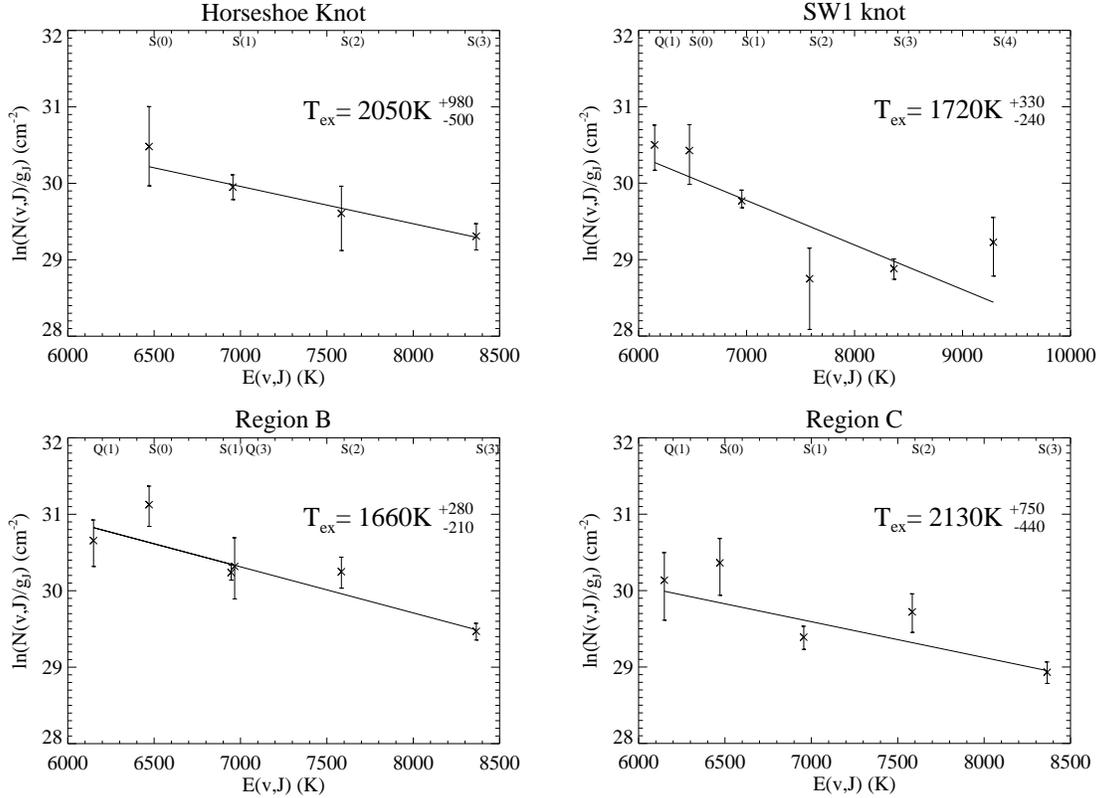}
\caption{Local thermodynamic equilibrium plots, where the inverse slope of the best fit line gives T$_{\rm ex}$. Errors on data points are at 1 $\sigma$ level, and error on T$_{\rm ex}$ is 1$\sigma$ error from weighted best fit. The labels refer to the $\vib{v}$=1$-$0\,S and Q transitions. $\vib{v}$=1$-$0\,S(1) and $\vib{v}$=1$-$0\,Q(3) transitions have been offset in region B for clarity.}
\label{temp}
\end{figure*}

The LTE plots (Fig. \ref{temp}) show the horseshoe knot spectrum fits a model of thermalised gas with an ortho-to-para ratio of three. 
The LTE plots of regions B and C show a clear misalignment of the ortho (odd J) and para (even J) lines which indicate a lower ortho-to-para ratio. 
The $\vib{v}$=1$-$0 S(4) line (and possibly $\vib{v}$=1$-$0 S(2)) are much narrower than the other molecular lines from the SW1 region, possibly affected by atmospheric sky subtraction (see Fig. 2).The S(0) line appears to agree with a model of thermalised gas with an ortho-to-para ratio of three, whilst the S(2) and S(4) lines do not.

Molecular hydrogen is formed with an ortho-to-para ratio of three. No radiative process can induce conversions between the ortho and para lines, hence the ratio is conserved. There are three known methods in which ortho-to-para conversions can occur: by collisions with protons in cold (T$<$100K) gas, or collisions with  hydrogen atoms in warmer gas (T$<$300K). Finally, if the hydrogen molecule stays on a grain long enough it can evaporate with an ortho-to-para ratio appropriate to the grain temperature.
In regions where shock excitation is the primary excitation mechanism, the rapid collisions cause the ortho-to-para ratio to take on a value close to three \citep{Smith}. However photodissociation regions (PDRs) often exhibit lower ortho-to-para ratios, typically 1.5-2.2. \citet{sternberg} provides models of far-UV (FUV) pumping in PDRs with optically thick H$_{2}$ ($\sim$10$^{14}$cm$^{-2}$ for UV) which explain how the observed ortho-to-para ratio will tend to 1.7 whilst the true ortho-to-para ratio remains at three due to lower rates of FUV pumping of molecular hydrogen in the optically thicker ortho states compared to the para states. In these models, the gas density must remain below 5$\times$10$^{4}$cm$^{-3}$ and be relatively cool (T$<$1000K) in order for collisional excitation to be negligible relative to FUV pumping, and collisional de-excitation to be negligible compared to spontaneous radiative decay, otherwise the ortho-to-para ratio would tend to the LTE value of three.

We follow \citet{DraineBert} definition of the ortho-to-para ratio, $\gamma$
\begin{equation}\nonumber
 \gamma=\frac{3a_{\rm ortho}}{a_{\rm para}}
\end{equation}
This ratio can be determined from a single pair of one ortho and one para line \citep{Puxley} through
\begin{equation}\nonumber
 \gamma=\frac{I_{o}}{I_{p}} \frac{A_{p}(2J_{p}+1)\lambda_{o}}{A_{o}(2J_{o}+1)\lambda_{p}}e^{\Delta E_{po}/{\rm k_{b}} T_{{\rm ex}}}
\end{equation}
where $\lambda$ is the wavelength, $J$ is the rotational quantum number of the upper state of the transition and $\Delta E_{po}$ is the energy difference between the upper energy level of the para and ortho line. Ortho/para lines are denoted by a o/p subscript.

T$_{{\rm ex}}$ and ortho-to para ratios are given in Table \ref{tempden} for regions B, C, SW1, and the horseshoe knot, where three or more H$_{2}$ lines are detected. The excitation temperature of the molecular hydrogen gas in the centre of NGC 1275 is given by \citet{Wilman} to be 1620\,K, whereas \citet{Krabbe} measure it to be 1450$\pm$250\,K (from the $\vib{v}=1$ transitions). T$_{{\rm ex}}$ of the filaments are consistent with these values for the gas in the core of the nebula, although tend to be slightly higher.

Although there are large errors, the ortho-to-para ratios suggests that regions B and C may also be non-thermally excited, possibly by FUV pumping in a PDR.

Assuming the upper levels are also populated in a Boltzmann distribution at the same excitation temperature, the total H$_{2}$ column density is given by
\begin{equation}\nonumber
N_{{\rm Total}}=\frac{N(\vib{v}, J)Z(T)}{g_{J}e^{-E(\vib{v}, J)/{\rm k_{b}}T_{{\rm ex}}}}
\end{equation}
where $Z(T)$ is the partition function. This expression assumes all the molecular hydrogen is collisionally excited at the same temperature and all the emission is visible. There are many scenarios in which this may not be the case, such as the X-ray dissociated regions modelled by \citet{Maloney}. In addition the line surface brightness listed in Table \ref{li} are only lower limits, therefore these column densities can be considered lower limits also. 

Column densities  are listed for each region  with three or more strong H$_{2}$ lines in Table \ref{tempden}. If T$_{{\rm ex}}$ of the upper levels is greater than in the lower levels, the column density will decrease from these values. All column densities are low (2$-$14 $\times$10$^{15}$cm$^{-2}$) which has implications for the structure of the filaments. For the gas to be thermalised the total density must be greater than 10$^{4}$\,cm$^{-3}$. The H$\alpha$ filaments are estimated to have a density of $\sim$10$^{2}$\,cm$^{-3}$ assuming they are in pressure equilibrium with the surrounding X-ray emitting gas. The filaments must thus have an undetected denser and cooler (1600-2200\,K) atomic component that mixes and shares energy with the molecular hydrogen. Alternatively, the H$_{2}$ exists in clumps, much smaller than a parsec in diameter, strung through the H$\alpha$ filaments. This has parallels with suggestions of \citet{Fabian2003} that the atomic hydrogen has a similar structure, on the basis of the H$\alpha$ emission and pressure balance. Consistently \citet{Wilman} argue that the H$_{2}$ emission near the nucleus of NGC 1275 must come from a population of dense self-gravitating clouds based on H$_{2}$ line ratios.
\begin{table}
\centering
\begin{tabular}{||l|c|c||c|}\hline
Region&Excitation & ortho-to-para& H$_{2}$ Column Density \\
&Temperature (K)&ratio ($\gamma$)&($10^{15}$\,cm$^{-2}$)\\ \hline
Horseshoe knot&2050$\pm^{980}_{500}$&2.8$\pm^{2.5}_{1.3}$ &  7.7 $\pm$2.0\\\hline
SW1 knot&1720$\pm^{330}_{240}$ &3.0$\pm^{3.2}_{1.2}$&10.1 $\pm$1.6\\\hline
Region B&1660$\pm^{280}_{210}$ &1.9$\pm^{0.8}_{0.5}$& 13.5$\pm$3.1\\\hline
Region C &2130$\pm^{750}_{440}$ &1.7$\pm^{1.0}_{0.6}$&2.5 $\pm$1.0\\ \hline
\end{tabular}
\caption{\label{tempden} Excitation temperature, ortho-to-para ratios and total line-emitting molecular hydrogen column density of all regions observed which revealed more than two molecular hydrogen lines. Errors correspond to the 1$\sigma$ level.}
\end{table}

\subsection{Comparison of Molecular and Atomic Hydrogen Emission}
Comparing our results with the H$\alpha$ distribution of \citet{conselice} (reproduced partly here in Fig. \ref{perreg}), we find the molecular hydrogen emission is clearly associated with the atomic emission, suggesting a common excitation mechanism.

Molecular to atomic line ratios are commonly used to distinguish between various excitation mechanisms. 3$\sigma$ upper limits of Br$\gamma$ were taken from our spectra and estimates of the H$\alpha$ emission from Fig. 5 of \citet{conselice} are used in conjunction with the strong $\vib{v}$=1$-$0 S(1) line to constrain excitation models.  

\begin{figure}
\centering
\includegraphics[width=1\columnwidth]{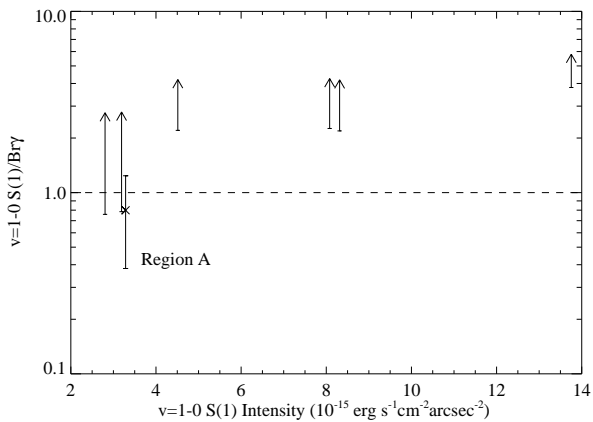}
\caption{The line intensity ratio $\vib{v}$=1$-$0 S(1)/Br$\gamma$ plotted against $\vib{v}$=1$-$0 S(1) intensity. $\vib{v}$=1$-$0 S(1) line has 1$\sigma$ errors, whilst Br$\gamma$ line is a 3$\sigma$ upper limit. Region A is labelled. Active galaxies and starforming regions have a ratio $\leq$ 1 (below the dashed line). The weakness of the Br$\gamma$ line rules out stellar UV as a possible excitation mechanism}
\label{s1tobrg}
\end{figure}

Active galaxies (starburst, Seyfert and ultra luminous IR galaxies) and star forming regions have the ratio $\vib{v}$=1$-$0 S(1)/Br$\gamma\leq1$ \citep{Jaffe2001} whereas most of the filament regions observed here have $\vib{v}$=1$-$0 S(1)/Br$\gamma\geq$ 1 (Fig. \ref{s1tobrg}), which strongly argues against a stellar UV excitation model. The exception is region A, which has a spectrum with $\vib{v}$=1$-$0 S(1)/Br$\gamma \le$1, although the molecular and Br$\gamma$ emission may have different origins as Fig.\ref{vel} suggests. 
All regions observed, except region A, do not show evidence for stars in the continuum images \citep{conselice} and regions further from the galaxy (SW1, SW2 and the horseshoe knot) do not show a continuum in their K-band spectra (Fig. \ref{Spectra}). It therefore seems unlikely that these regions can be excited by UV photons from young, hot stellar populations. However Ly$\alpha$ emission and  UV continuum has been found associated with the optical filaments \citep{Fabian1984,Nord} that may not have originated from stars.

\begin{figure}
\centering
\includegraphics[width=1\columnwidth]{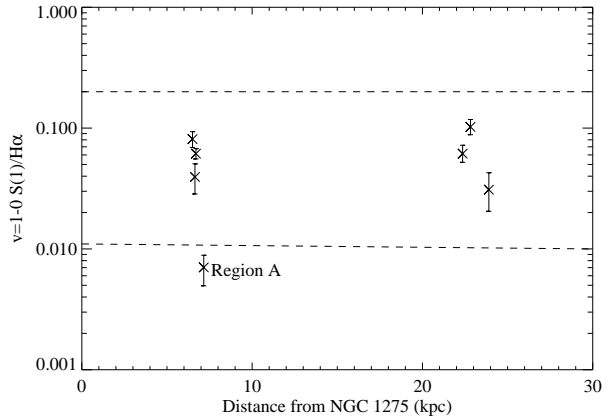}
\caption{The line intensity ratio $\vib{v}$=1$-$0 S(1)/H$\alpha$ as a function of distance from the nucleus of NGC 1275. The region bounded by the dashed lines corresponds to the range of $\vib{v}$=1$-$0 S(1)/H$\alpha$ ratios in other nuclear regions of central cluster galaxies \citep{Wilman}. As there is no correlation with radius, the central AGN can not be exciting the molecular hydrogen}
\label{s1toHa}
\end{figure}
\begin{figure}
\centering
\includegraphics[width=1\columnwidth]{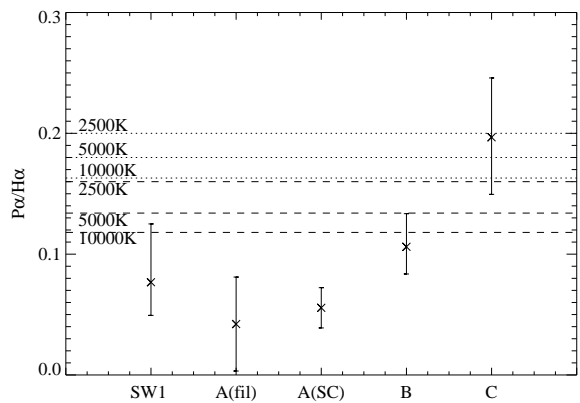}
\caption{The line intensity ratio P$\alpha$/H$\alpha$ for regions in which  P$\alpha$ could be measured or a reasonable estimate made. All errors follow from 1 $\sigma$ errors on the line intensities. The dotted lines give the expected ratio given Case A recombination with gas at the temperature written above the line. The dashed lines give the expected ratio given Case B recombination with gas at the temperature written below the line \citep{Osterbrock}. P$\alpha$ emission from region A is divided into the broad and narrow component, with 30 per cent of the detected H$\alpha$ emission assumed to have come from the filamentary material, A(fil), and 70 per cent from the Stellar Cluster, A(SC).}
\label{paha}
\end{figure}
The $\vib{v}$=1$-$0 S(1)/H$\alpha$ ratios for the outer filaments, shown in Fig. \ref{s1toHa}, generally lie between 0.02-0.1, similar to ratios from the centre of other dominant cluster galaxies. In the central region of NGC 1275 \citet{Donahue} show the $\vib{v}$=1$-$0 S(1)/H$\alpha$ ratio decrease from 0.19 to below 0.01 out to a radius of 4 arcsec. Following \citet{Shields2} estimate of the ratio of H$\alpha$ emission in the filament compared to the stellar cluster of region A, only 30 per cent of the measured H$\alpha$ surface brightness has been taken into account for region A whereas all the $\vib{v}$=1$-$0 S(1) emission has been taken into account. Even with this estimate, the ratio of region A falls below the other regions. 
Fig. \ref{s1toHa} shows no correlation with distance from the central galaxy as would be expected if the active nucleus of NGC 1275 enhanced the  H{\sc ii} or H$_{2}$ emission. We infer that the AGN can only play a minor role in exciting the filaments \citep{RodAndy}.

Possible excitation mechanisms that could produce these ratios have been discussed in a number of papers. \citet{Donahue} suggests X-ray heating of thick columns of molecular hydrogen, conduction from the ICM or mixing layers. These ratios could also be obtained from shocks according to the models of \citet{Hollenbach}, whilst predictions of CLOUDY models, of dense self-gravitating clouds with n$_{{\rm T}}=10^{5}$\,cm$^{-3}$ exposed to UV radiation also produce these ratios \citep{Wilman}.

The H$\alpha$ intensity of the H$_{2}$ regions was estimated using aperture photometry on the \citet{conselice} image (Fig \ref{perreg}) over the regions were the slits crossed. In doing so errors may be introduced by, for example, picking slightly different regions and different calibration methods. In addition no correction has been made to remove the [N{\sc ii}] line flux from the image or take into consideration any reddening which could have a large effect on this ratio. As it is difficult to assign an error to this, we can instead measure the P$\alpha$/H$\alpha$ ratios and compare them with predicted recombination ratios. Fig \ref{paha} shows the P$\alpha$/H$\alpha$ ratios for regions SW1, A, B and C together with various recombination ratios given by \citet{Osterbrock}. Although there is a large scatter, regions SW1, B and C all agree within error to Case B recombination ratios. The  P$\alpha$/H$\alpha$ ratio for both the broad filamentary emission and the narrow stellar cluster emission of region A lies well below any expected recombination ratio. Therefore it is possible that the $\vib{v}$=1$-$0 S(1)/H$\alpha$ ratio of region A is badly compromised by an inaccurate H$\alpha$ surface brightness estimate as well as an uncertain $\vib{v}$=1$-$0 S(1) line surface brightness.

\subsection{Pressure Balance}
The X-ray emitting gas surrounding the molecular hydrogen has a number density of n$\sim$0.06\,cm$^{-3}$, and a temperature of approximately 3$\times 10^{7}$\,K, giving a thermal pressure of $\sim 1.8 \times 10^{6}$\,cm$^{-3}$K \citep{Sanders}. The pressure of the outer regions of the optical line-emitting gas has been estimated from the [S{\sc ii}] doublet \citep{Heckman} and the gas is found to be approximately in pressure equilibrium with the surrounding X-ray medium. As the molecular hydrogen line ratios imply the emission is partially if not predominately thermal, with T$\sim$1600-2200\,K, the total density from at least part of the molecular line-emitting region must be greater than the critical density above which thermal emission dominates, n$\ge$10$^{4}$\,cm$^{-3}$. This implies the molecular line-emitting regions must be over-pressurized by at least 10 times that of its surrounding medium. This echoes the pressure problem raised by  \citet{Wilman} and \citet{Jaffe2001} who postulated the need for a dynamic, non-isobaric model in order to explain both the ionized and molecular emission.

\section{Conclusions}
We have found molecular hydrogen emission from the same regions as the H$\alpha$ emission in the outer filaments, up to 24\,kpc from the nucleus of the central cluster galaxy, NGC 1275. As H$_{2}$ quickly dissociates above temperatures of 4000\,K, this finding proves that a cooler component is intimately associated with the extended optical/UV line-emitting filaments. 
The spectra are consistent with predominately thermal emission, whilst the ortho-to-para ratios of regions B and C suggest an additional non-thermal excitation component possibly from UV photons in the Lyman-Werner band. The high H$_{2}$/Br$\gamma$ ratios suggest that UV radiation from hot, young stars is unlikely, however there is Ly$\alpha$ emission and UV continuum associated with the optical filaments and X-ray irradiation of the molecular hydrogen by the surrounding ICM may be able generate enough photons in Ly$\alpha$ and in the Lyman-Werner band through non-thermal electron collisions with hydrogen atoms \citep{Maloney}. 

The measured excitation temperatures lie between 1600-2200\,K and the column densities of the warm excited molecular hydrogen in these regions are 2$-$14$\times$10$^{15}$\,cm$^{-2}$. It is possible that there exists an even cooler component of molecular gas within these filaments which will be revealed only by searches for the $\vib{v}=(0-0)$ pure rotational lines of hydrogen in the mid-IR, or CO emission in the mm waveband. As the emission is thermal, at least some of the emitting gas has a density greater than 10$^{4}$cm$^{-3}$. Comparing the minimum pressure of the molecular line-emitting gas to that of the X-ray gas shows the molecular gas must be over-pressurized by at least an order of magnitude.

The presence of this molecular gas so far out in the ICM poses similar challenges as the presence of the optical line-emitting gas. Either the gas has formed in situ (having cooled from the ICM and is falling onto the galaxy in a condensing phase), or the gas has travelled from a much cooler region, abundant in molecular hydrogen, to where it has been detected in the ICM. 

A large reservoir of molecular hydrogen is known to lie in the nucleus of NGC 1275 \citep{Jaffe2001,Edge,Donahue,Krabbe}. The origin of this gas is still debated, but it may have come from a previous merger as suggested by \citet{Braine} or be accumulated through cooling of the ICM. Theories have been proposed suggesting that the buoyant old radio bubbles, visible in X-ray images, drag gas from the central regions out into the ICM as they rise \citep{Bohringerm87,Churazov,Fabian2003}. In these models the filaments act as streamlines tracing the path of the cooler gas as it moves and is heated by the ICM. 

The horseshoe-shaped filament and its mirror image seen to the West (see Fig. \ref{perreg}) in the NW, appears just below an old radio bubble with a flow pattern similar to that of water underneath a rising air bubble (see \citealt{Fabian2003} for discussion). It is possible that the molecular hydrogen originated from the core of NGC 1275 and it is currently being heated as it is dragged into the hot ICM behind these buoyant bubbles. However not all optical emission-line filaments with molecular hydrogen appear to follow visible radio bubbles; in particular the radial filaments and molecular knots in the East do not point toward any known radio bubble. These filaments could have been drawn out by a bubble whose radio emission is now too  faint to detect, although that would suggest that the molecular hydrogen exists in the hostile environment of the ICM for about $\sim$10$^7-$10$^8$ years, thus having clear implications on the rate of conduction within the ICM. High velocity resolution spectroscopy of these filaments would enable mapping of any flow pattern and thus probe further the origin of this cool gas.

\section*{Acknowledgements}
NAH and RMJ acknowledge support from PPARC and ACF and CSC thank the Royal Society for support. We thank the UKIRT support staff for all their help. The United Kingdom Infrared Telescope is operated by the Joint Astronomy Centre on behalf of the U.K. Particle Physics and Astronomy Research Council.

\bibliographystyle{mnras}
\bibliography{mn-jour,ME982rv}

\end{document}